\documentclass[12pt,a4paper,oneside]{article}
\usepackage{graphicx}
\usepackage{color}
\usepackage{epsfig}
\usepackage{multirow}
\usepackage[english]{babel}
\usepackage{amsmath}
\usepackage{amssymb}
\usepackage[latin1]{inputenc}
\usepackage{indentfirst}
\usepackage{newlfont}
\usepackage{latexsym}
\usepackage[latin1]{inputenc}
\usepackage[a4paper,top=4cm,left=4cm]{geometry}

\def\a{\alpha}
\def\b{\beta}

\def\g{\gamma}

\newtheorem{theorem}{Theorem}
\newtheorem{lemma}[theorem]{Lemma}
\newtheorem{proposition}[theorem]{Proposition}
\newtheorem{corollary}[theorem]{Corollary}
\newtheorem{definition}[theorem]{Definition}

\title{A new approach to separation of variables for the Clebsch integrable system. Part I: Reduction to quadratures  }
\author{ Y. Fedorov$^{1}$, F. Magri$^{2}$, T. Skrypnyk$^{3,4}$ \\
{\small $^{1}$ Polytechnic University of Catalonia, Barcelona, Spain}\\
{\small $^{2}$ Dipartimento di Matematica e Applicazioni- Universit\'{a} di Milano Bicocca, Milano, Italia}\\
{\small $^3$ Universit\'{a} degli Studi di Torino, via Carlo Alberto 10, 10123, Torino, Italia }\\
{\small $^{4}$ Bogolyubov Institute for Theoretical Physics, Metrolohichna str.14-b, 03115, Kiev, Ukraine}\\
{\footnotesize yuri.fedorov@upc.edu, franco.magri$@$unimib.it, taras.skrypnyk@unimib.it}}
\date{}
\begin{document}
\maketitle
\begin{abstract}
The paper describes a new concept of separation of variables with a concrete application to the Clebsch integrable case of the Kirchhoff equations. There are two principal novelties:  
The first is that the separating coordinates are constructed (not guessed) by solving the Kowalewski separability conditions. 
The second is that the solutions of the equations of motion are written in terms of theta-functions by means of 
a generalization of the standard Jacobi inversion problem of algebraic geometry. These two novelties are dealt with  
in two separate parts of the paper. Part I explains the Kowalewski separability conditions and their implementation 
to the Clebsch case. It is shown that the new separating coordinates lead to quadratures involving Abelian 
differentials on two different non-hyperelliptic curves (of genus higher than the dimension of the invariant tori). In Part II these quadratures are interpreted as a new generalization of the standard Abel--Jacobi map, and a procedure of its inversion in terms of theta-functions is worked out. The theta-function solution is different from that found long time ago by F. K\"otter, since 
the theta-functions used in this paper have different period matrix.
\end{abstract}

\section{Introduction}

The famous Clebsch integrable case of the Kirchhoff equations \cite{Clebsch} belongs to an ample class of classical and modern 
algebraic integrable systems, for which finding a separation of variables is an especially difficult problem.
The class also includes, for instance, the Euler-Frahm top on the algebra $so(4)$, integrated by Schottky \cite{Schot}, the Kowalewski top, the Henon-H\'eiles system, and the geodesic flow on $SO(4)$ \cite{AvMVh, avm82}. For the general case of motion, 
the Clebsch case has been first solved in theta-functions by F. K\"otter \cite{kot892}, who devised 
a reduction to quadratures involving a pair of points on a genus 2 hyperelliptic curve (the so called K\"otter's curve). The coordinates of the points 
give the separating coordinates, and the quadratures themselves have the form of a standard Abel--Jacobi map. Later on, the theta-function solutions for 
the Clebsch case and the Euler--Frahm top have been reconstructed in a series of publications, including \cite{Bel,ZC}, where the method of  Lax 
representation and of the related  Baker--Akhiezer functions \cite{Skl} was used instead of separation of variables. Furthermore, the algebraic geometric structure of the complex invariant tori, and the linearization of the flow from the standpoint of the theory of Abelian varieties, have been described in detail in \cite{avm84, hai83}.

\medskip
The present paper aims to revise the approach based on the method of separation of variables. Our approach is unconventional both for the technique used to find the separating coordinates, and for the technique used to solve  the equations of motion in terms of theta-functions. The first difference is that to find the separating coordinates we utilize the technique of  the Kowalewski separability 
conditions  \cite{MaKow1, MaKow2, MaKow3}. This means that the separating coordinates are systematically obtained 
(not guessed) by solving the above conditions in each specific example. The second difference is that our quadratures involve sums of Abelian integrals on two different non-hyperelliptic curves $C$ and $K$ of genus 3 (while the quadratures of K\"otter involve a single hyperelliptic curve 
of genus 2). They lead to a new generalization of the Abel-Jacobi map. This unexpected feature does not forbid, however, to invert the map and to write the solutions of the equations of motion in terms of theta-functions. These novelties are tersely discussed below.

\medskip
The Kirchhoff equations describing the motion of a rigid body in an ideal fluid can be written in the form
\begin{gather*}
\dot{\vec{S}}=\vec{S} \times \frac{\partial H}{\partial \vec{S}}+ \vec{T} \times \frac{\partial H}{\partial \vec{T}}\, ,\qquad
\dot{\vec{T}}=\vec{T} \times \frac{\partial H}{\partial \vec{S}} \,,
\end{gather*}
where $\vec S $ and $\vec T $ are vectors in the three-dimensional Euclidean space $\mathbb{E}_{3}$, and $H$ is a scalar function.  
These equations can be interpreted as a Hamiltonian vector field $X_{H}$ on the dual to the Lie algebra of the group of motions of $\mathbb{E}_{3}$. The Clebsch integrable case corresponds to the choice 
\begin{equation*}
H= \sum\limits_{\a=1}^3m_{\a}S^2_{\a}+\sum_{\a=1}^3 n_{\a}T^2_{\a} ,
\end{equation*}
where the coefficients satisfy the constraint
\begin{equation*}
\frac{n_{1}-n_{2}}{m_{3}} + \frac{n_{2}-n_{3}}{m_{1}} +\frac{n_{3}-n_{1}}{m_{2}} = 0 \, .
\end{equation*}
It is convenient to represent these coefficients in the form
\begin{gather*}
m_{\a}= \sigma+\tau j_{\alpha} \quad
n_{\a}= \sigma (j_{\beta}+j_{\gamma})+\tau j_{\beta}j_{\gamma} , \qquad \sigma, \tau \in {\mathbb R}, \qquad (\alpha,\beta,\gamma)=(1,2,3) ,
\end{gather*}
$j_1, j_2, j_3$ being arbitrary parameters;  
and to regard $H$ as the linear pencil $ H = \sigma H_{p} + \tau H_{s} $ generated by the Hamiltonians
\begin{gather} \label{Hps}
H_{p} = \sum\limits_{\a=1}^3 S_{\a}^{2} + ( j_{\b} + j_{\g} ) T_{\a}^{2}, \quad
H_{s} = \sum\limits_{\a=1}^3  j_{\a} S_{\a}^{2} +  j_{\b}  j_{\g}  T_{\a}^{2} .
\end{gather}
In the sequel we assume that $j_1, j_2, j_3$ are distinct. 

By our definition, the Clebsch system 
is the pair of Hamiltonian vector fields $X_{p}$ and $X_{s}$ associated with the above Hamiltonians and given by 
\begin{align}\label{ClebEq}
\begin{split}
X_{p}(S_{\a}) &= (j_{\g}-j_{\b})T_{\b}T_{\g},\\
X_{p}(T_{\a}) &= S_{\b}T_{\g}-S_{\g}T_{\b}
\end{split}
\end{align}
and, respectively,
\begin{align}\label{ClebEq'}
\begin{split}
X_{s}(S_{\a}) &= j_{\a}(j_{\g}-j_{\b})T_{\b}T_{\g}+ (j_{\b}-j_{\g})S_{\b}S_{\g},\\
X_{s}(T_{\a}) &= j_{\b}S_{\b}T_{\g}-j_{\g}S_{\g}T_{\b} .
\end{split}
\end{align}
It is known that the 
vector fields $X_{p}$ and $X_{s}$ commute, and that they possess four integrals of motion, namely the above Hamiltonians 
$H_{p}$ and $H_{s}$, as well as the functions
\begin{gather}\label{Casim}
C_1=\sum\limits_{\a=1}^3T_{\a}S_{\a}, \quad \qquad
K_1=\sum\limits_{\a=1}^3T^2_{\a}.
\end{gather}

The procedure for explicitly solving these differential equations  
 consists of two stages. 
First, we introduce a suitable set of separating coordinates. Namely, in the complex projective space ${\mathbb P}^5$ with homogeneous coordinates $(A_{\a}, B_{\a})$, $\a=1,2,3$, consider the intersection of four quadrics given by equations
\begin{equation}\label{quadrics}
\begin{aligned}  \sum\limits_{\a=1}^3 ( j_{\b} + j_{\g} ) B_{\a}^{2} + A_{\a}^{2} &= 0 \,, \quad   
\sum\limits_{\a=1}^3  j_{\b}  j_{\g}  B_{\a}^{2} + j_\a A_{\a}^{2} = 0 \, ,  \\
\sum\limits_{\a=1}^3 A_{\a} B_{\a} & = 0\,, \quad  
\sum\limits_{\a=1}^3 B_{\a}^{2} = 0\,,    \qquad (\a,\b,\g)=(1,2,3).   
\end{aligned}
\end{equation}
Since the equations are independent, the intersection is a curve ${\cal E}\subset {\mathbb P}^5$. Its intersection with the hyperplane
\begin{equation}
\sum\limits_{\a=1}^3 A_{\a} T_{\a} + B_{\a} S_{\a} = 0 \label{plane}
\end{equation}
gives eight points, which move along ${\cal E}$ while the coordinates $S_{\a}, T_{\a}$ vary in time with the fields $X_{p}$ and $X_{s}$. 
With each point we associate the complex parameter
\begin{equation}
v = - \frac{\sum_{\a=1}^{3}  j_{\a} A_{\a}^{2}}{\sum_{\a=1}^{3}  A_{\a}^{2}} .  \label{v}
\end{equation}
Let $v_{p}$ and $v_{s}$ denote its derivatives along the vector fields $X_p$ and $ X_s$ respectively. Then each intersection point will give us the pair of coordinates 
\begin{gather}
x_{1} = v \, ,\qquad
x_{2} = - \frac{v_{s}}{v_{p}  } .
\end{gather}
The eight pairs of $( x_1, x_2 )$ constructed in this way have the remarkable property to verify the same system of differential equations. Namely, let  
\begin{equation*}
g(x) = K_1 x^2+ H_p x + H_s \, ,
\end{equation*}
and let $dx_{1}= x_{1p} d t _{p} + x_{1s} d t_{s}$, $dx_{2}= x_{2p} d t _{p} + x_{2s} d t_{s}$ be the differentials of $x_1,x_2$, where $(x_{1p}, x_{2p})$ and $(x_{1s}, x_{2s})$ denote their derivatives with respect to the vector fields $X_{p}$ and $X_{s}$, and $t_{p}$ and $t_{s}$ are the time parameters along these fields. Then the system reads 
\begin{align}\label{Ab}
\begin{split}
\frac{d x_{1}}{w( w^2 - g(x_1) ) }+ \frac{\sqrt{2} \, d x_{2 }}{W (W^2 - g(x_2))} &= \frac{i}{ C_1} d t_{s} \, ,\\
\frac{x_1 \,d x_1 }{w( w^2 - g(x_1) )  }+  \frac{\sqrt{2}\, x_2\, d x_{2 }}{W (W^2 - g(x_2) ) }&=\frac{i}{ C_1} d t_{p}\, , \qquad 
i =\sqrt{-1}\,. 
\end{split}
\end{align}
Here the the pairs of the variables $(x_1,w)$ and $(x_2,W)$ satisfy equations of algebraic curves 
\begin{align}
\begin{split}
 &C : \qquad w^4 - 2 g(x_1)w^2 + g^2(x_1) - 4 C_1^2 (x_1+j_1)(x_1+j_2)(x_1+j_3) = 0 \,, \\ \label{curves}
 &K : \qquad W^4 -2 g(x_2) W^2 + 4 C_1^2 (x_2+j_1)(x_2+j_2)(x_2+j_3) =0\, . 
 \end{split}
\end{align}
These are the genus 3 non-hyperelliptic curves $C$ and $K$ mentioned above. In the integral form, the equations \eqref{Ab} give the quadratures 
   \begin{equation} \label{A-P_0}
 \int_{P_0}^P \begin{pmatrix} \omega_1 \\ \omega_2  \end{pmatrix} +   \int_{R_0}^R \begin{pmatrix} \bar \omega_1 \\ \bar \omega_2  \end{pmatrix} =
 \begin{pmatrix} t_s \\ t_p  \end{pmatrix}, \qquad P=(x_1,w)\in C,\quad R=(x_2, W)\in K, 
\end{equation}
where $(\omega_1, \omega_2)$ and $(\bar\omega_1, \bar\omega_2)$ are certain holomorphic differentials on the curves $C$ and $K$ respectively, and $P_0, R_0$ are some base points on them. 

The second stage of our procedure is the inversion of the quadratures. The facts that the genus of 
the curve is higher than the dimension of the invariant tori and that the periods of the Abelian integrals on the two curves are distinct, raise a 
natural question about the invertibility of \eqref{A-P_0} in terms of meromorphic functions of the complex times $t_p, t_s$.  Nevertheless, by using the results of \cite{avm84, hai83, Pant}, we show that the quadratures describe a well-defined map $\cal P$ from $C\times K$ to the two-dimensional Prym subvariety of the Jacobian of $K$. We call it the Abel--Prym map. To our best knowledge, such kind of maps did not appear   
before, neither in the classical nor in the modern literature. Moreover, in contrast to the standard Abel--Jacobi map, which is one-to-one almost 
everywhere, under the Abel--Prym map each point of the Prym variety has 8 preimages $(P_1,R_1),\dots, (P_8,R_8)$ on $C\times K$. They can 
be found as zeros of a higher order theta-function $\Xi^*$ which is quasiperiodic on the Prym variety, that is, by applying a generalization of the Riemann vanishing theorem, which we formulate and prove.  
 
As a result, any symmetric functions of the coordinates of $(R_1,\dots, R_8)$ (or of $(P_1,\dots, P_8)$) can be written in terms of $\Xi^*$ and its 
derivatives.  On the other hand, the coordinates of these preimages coincide with the eight sets of separating variables found initially from the point $(S_{\a}, T_{\a})$ in the phase space.    
The variables $(S_{\a}, T_{\a})$  can be written in terms of symmetric functions of the eight pairs 
of separating coordinates. Combining these results enables one to obtain theta-function solutions for the Clebsch integrable case.

\medskip
Part I of the paper deals with the first stage: construction of separation variables and reduction to quadratures. 
Section 2 introduces the Kowalewski separability conditions in the context of the theory of Poisson pencils and bi-Hamiltonian geometry. In Section 3 these conditions are used to compute the separating coordinates for the Clebsch system. These coordinates are studied in Section 4, where the quadrature formula shown 
before is explicitly worked out. Finally, Section 5 contains  short comments on potential extensions of the technique developed in the paper. 
The appendix contains the proof of the Main Proposition stated in Section 3.
\medskip

Part II deals with the second stage: analysis of the main properties of the Abel-Prym map generated by the quadratures \eqref{A-P_0} and  
the algorithm of its inversion in terms of theta-functions of higher order, which lead to a new theta-function solution of the Clebsch system. 
This part consists of three sections.

\medskip
The two parts are kept separate because each of them has its own intrinsic interest, independent of the example of the Clebsch system.
They correspond to different but complementary ways of approaching the study of separable systems: the differential-geometric 
viewpoint of the theory of Poisson manifolds, and the algebraic-geometric viewpoint of the theory of Abelian varieties. Also the languages used in the two parts are 
quite different. These peculiarities suggest to deal with them in  separate form. Notwithstanding, the two parts must be considered together, 
because both of them are necessary to provide a full picture of the behavior of the solutions of the Clebsch equations of motion.

\section{The Kowalewski separability conditions and the theory of Poisson pencils}

In this section we introduce a particular class of integrable Hamiltonian vector fields and we investigate their separability properties. The class contains the Clebsch system. The initial point of the investigation is the concept of Poisson pencil. The final point is a particular form of the Kowalewski separability conditions. They allow to recover, quite rapidly, the particular solution of the Clebsch system obtained by Weber in 1878 \cite{web878}. The analysis of this elementary example will serve as guideline for the general study of the Clebsch system carried out in the next section.

\medskip
Let $M$ be a smooth manifold of dimension $m$, and $P : T^{*}M \rightarrow TM$,  $Q : T^{*}M \rightarrow TM$ be a pair of bivectors defined on $M$ (that is, second-order skew-symmetric contravariant tensor fields). Assume that $P$ and $Q$ verify the  Schouten conditions
\begin{equation*}
[ P , P ]_{Sch} = 0\,, \quad [ P , Q ]_{Sch} = 0\, , \quad [ Q , Q ]_{Sch} = 0 \,,
\end{equation*}
where $ [\cdot \  ,\ \cdot ]_{Sch} $ denotes the Schouten bracket on the exterior algebra of multivectors. Then $M$ is called a bihamiltonian manifold, and the pencil of bivectors 
$Q - \lambda P$, $\lambda$ being a real or complex parameter, is called a Poisson pencil. The interest for these pencils is due to their ability of generating integrable Hamiltonian systems through their polynomial  Casimir functions \cite{GelZak}. Let
\begin{equation*}
g(\lambda) = K_1 \lambda^{q} +K_2 \lambda^{q-1} + \cdots + K_{q+1}
\end{equation*}
be a polynomial with coefficients in the ring of differentiable functions on $M$, and let
\begin{equation*}
dg(\lambda) = dK_1 \lambda^{q} + dK_2 \lambda^{q-1} + \cdots + dK_{q+1}
\end{equation*}
be its differential. The function $g(\lambda)$ is said to be a polynomial Casimir function of the pencil $Q - \lambda P $ if
\begin{equation*}
(Q - \lambda P )dg(\lambda) = 0
\end{equation*}
for any value of $\lambda$. In this case, the first coefficient $K_1$ is a Casimir function of $P$, the last coefficient $K_{q+1}$ is a Casimir function of $Q$, and the intermediate coefficients satisfy the recursive relations
\begin{equation*}
Q dK_{a} = P dK_{a+1} .
\end{equation*}
There are $q$ of such relations. Each of them defines a vector field $X_{a} = Q dK_{a} = P dK_{a+1} $, for $a= 1,\dots, q$. Thus a polynomial Casimir function of degree $q$ endows $M$ with $q+1$ distinguished functions $(K_1,K_2,\dots , K_{q+1} )$ and with $q$ distinguished vector fields $(X_1, X_2, \dots , X_q) $. Assume that $Q -\lambda P$ has $n$ polynomial Casimir functions of degrees $(q_{1}, q_{2}, \dots , q_{n})$, and suppose that the dimension $m$ of the manifold,  the number $n$ of polynomial Casimir functions, and the  sum $p= q_1+\cdots + q_{n}$ of the degrees of the polynomial Casimir functions fulfill the relation $m= n+2p $. Then, it is a well known result of the theory of Poisson pencils that each vector field defined by the pencil is an integrable Hamiltonian system in the sense of Liouville. Let us see how this works for the Clebsch system.

\medskip
\noindent
\begin{paragraph}{Example 1.} On two copies of the Euclidean space $\mathbb{E}_{3}$, equipped with coordinates $(S_{\a}, T_{\a} )$, consider four bivectors:
\begin{align*}
P_{0} &= \sum\limits_{\a} S_{\a} \frac{\partial}{\partial{T_{\b}}} \wedge \frac{\partial}{\partial{T_{\g}}} \, , \\
P_{\a} &= S_{\a} \frac{\partial}{\partial{S_{\b}}} \wedge \frac{\partial}{\partial{S_{\g}}} + (  T_{\b}\frac{\partial}{\partial{S_{\g}}} - T_{\g}\frac{\partial}{\partial{S_{\b}}} ) \wedge \frac{\partial}{\partial{T_{\a}}}\,, \qquad (\a,\b,\g)=(1,2,3) . 
\end{align*}
Any pair of them verifies the Schouten conditions defining a Poisson pencil, and, therefore, any linear combination of $P_0, P_1, P_2, P_3$ with constant coefficients is again a Poisson bivector. Consider the bivectors
\begin{align*}
Q =P_{0} +j_1 P_{1} +j_2 P_{2} +j_3 P_{3}\, , \qquad  P = P_{1} + P_{2} + P_{3}.
\end{align*}
They endow $M$ with the structure of a bihamiltonian manifold. The corresponding Poisson pencil $Q - \lambda P $ has two polynomial Casimir functions. The first has degree zero,
\begin{equation*}
g_{1}(\lambda) = C_1 .
\end{equation*}
The second has degree two:
\begin{equation*}
g_{2}(\lambda) = K_1 \lambda^{2} + K_2 \lambda +  K_{3} .
\end{equation*}
Thus the Poisson pencil defines two vector fields $X_1$ and $X_2$. Since the relation $m= n+2p$ is verified, they form an integrable 
bihamiltonian hierarchy. Working with the coordinates $(S_{\a}, T_{\a} )$ one may check easily that the functions 
$(C_1, K_1, K_2, K_3)$ and the vector fields $(X_1, X_2$) coincide with the functions $(C_1, K_1, H_p, H_s)$ and the vector fields $(X_{p}, X_{s})$ of the previous section (possibly up to an irrelevant common constant multiplicative factor). Hence one may claim that the Clebsch system is 
the bihamiltonian hierarchy defined by the Poisson pencil discussed in this example. $\square$
\end{paragraph}

\medskip
Having defined the class of integrable systems we are interested in, the attention is now addressed to the problem of their separability. 
The objective is to select a class of Poisson pencils whose bihamiltonian hierarchy is separable in a system of  separating coordinates 
\emph{provided by the pencil itself}. To discuss this problem we have recourse to the theory of Kowalewski separability conditions, from which
we extract the following particular result. 

\begin{proposition} Let $X_{p}$ and $X_{s}$  be a pair of commuting vector fields on a manifold $M$ of dimension $m$, possessing $(m-2)$ independent integrals of motion $(H_1, \dots , H_{m-2})$. Let furthermore $h(x)=E x^{2} + F x + G $ be a quadratic polynomial whose coefficients satisfy the Kowalewski separability conditions
\begin{gather}\label{Kow}
E F_{s} - F E_{s} = E G_{p} - G E_{p}\,,  \quad
E G_{s} - G E_{s} = F G_{p} - G F_{p} ,
\end{gather}
where the symbols $E_{p}, F_{p},  G_{p}, E_{s}, F_{s}, G_{s}$ denote the derivatives of the coefficients of $h(x)$ along the vector fields 
$X_{p}$ and $X_{s}$ respectively. Then the roots $x_1$ and $x_2$ of the quadratic equation
\begin{equation}\label{SP}
E x^{2} + F x + G = 0 
\end{equation}
are separating coordinates for $X_{p}$ and $X_{s}$.
\end{proposition}

{\it Proof.} Let $\cal{F}$ be the 2-dimensional foliation spanned by the vector fields $X_p$ and $X_s$. The functions $H_1, \dots , H_{m-2}$ are constant on the leaves of the foliation. Let us add to these functions the roots $x_1$ and $x_2$ of the polynomial $E x^2 + F x + G $. Together they form a coordinate system on $M$ adapted to the foliation. In view of the Kowalewski conditions and the $Vi\grave{e}te$ formulas $x_1+x_2=-F/ E$ and $x_1x_2=G/E$, one easily finds that the roots $x_1$ and $x_2$  satisfy the differential equations
\begin{equation} \label{Kowdual}
x_{1s} + x_{2} x_{1p} = 0,  \qquad \qquad \qquad x_{2s} + x_{1} x_{2p} = 0 .
\end{equation}
Consider then the vector fields $X_s + x_1 X_p $ and $X_s + x_2 X_p  .$  The previous relations entail that these vector fields commute, and that $X_s + x_1 X_p $ is a multiple of $\partial/\partial{x_1}$, while 
$X_s + x_2 X_p $ is a multiple of $\partial/ \partial{x_2}$. Hence, there exist two functions $\psi _{1}$ and $\psi _{2}$ of the coordinates $( x_1, x_2, H_1, \dots , H_{m-2} )$ such that
\begin{align}
\begin{split}
\psi_1 \frac{\partial}{\partial{x_1}} &= X_s + x_1 X_p  \\  \label{Stack}
\psi_2 \frac{\partial}{\partial{x_2}} &= X_s + x_2 X_p  .
\end{split}
\end{align}
By  the commutativity of  $X_s + x_1 X_p $ and $X_s + x_2 X_p $, the function $\psi_1$ does not depend on $x_2$, whereas $\psi_2$ does not depend on $x_1$. This is a key property in the process of separation of variables.

To proceed further, assume that the multipliers $\psi_1$ and $\psi_2$ are algebraic functions of $x_1$ and $x_2$ respectively, that is they are rational on certain algebraic curves 
$C \subset {\mathbb C}^2 (x_1,w)$ and $K\subset {\mathbb C}^2 (x_2,W)$. Thus one may set $\psi_1= R_1(x_1, w)$ and $\psi_2= R_2(x_2, W)$. 
In terms of differential forms, the vector equations (\ref{Stack}) then read
\begin{align} 
\begin{split}
\frac{ d x_1}{R_1(x_1, w)} + \frac{ d x_2}{R_2(x_2, W)} & = d t_{s} \\   \label{Ab2}
\frac{x_1  d x_1}{R_1(x_1, w)} + \frac{x_2  d x_2}{R_2(x_2, W)} & = d t_p \, ,
\end{split}
\end{align}
where, as above, $t_{p}, t_{s}$ are the time parameters along $X_p, X_s$ respectively.
 
If the functions $R_1, R_2$ coincide, the above quadratures represent a kind of  Abel--Jacobi map (standard or generalized), which can be inverted in terms of the Riemann theta-functions or their generalization. 
However, if $R_1, R_2$ are different and the corresponding curves $C$ and $K$ are not birationally equivalent, the problem of inversion of \eqref{Ab2} may be more delicate or even not solvable: for the case of the Clebsch system it will be addressed in the second part of the paper. $\square$

\bigskip
To relate the theory of Poisson pencil with the Kowalewski separability conditions we recall that any Poisson bivector $P$ defines a derivation 
$d_{P}$ on the exterior algebra of multivectors. It is defined by $d_{P} = [ P , \ \cdot ]_{Sch} $, has degree 1, and verifies the cohomological 
condition $ d_{P}^{2} = 0 $. It  allows to define the special class of vector fields $Z$ for which $Z \wedge d_{P} Z =0$ on the whole
manifold $M$ or, at least, on a level surface of the Casimir functions of $P$. Here is an example of such a vector field. 

\medskip
\noindent
\begin{paragraph}{Example 2.} Consider again the Poisson pencil $Q- \lambda P$ of the previous example, and the vector field $Z = \sum_{\a}  T_{\a} \frac{\partial}{\partial{T_\a}}$. The Lie derivative of $P$ along $Z$ vanishes, while the Lie derivative of $Q$ is
\begin{equation*}
Lie_{Z}Q = -2 \sum_{\a=1}^{3} S_\a \frac{\partial}{\partial{T_\b}} \wedge \frac{\partial}{\partial{T_\g}}. 
\end{equation*} 
Since  $d_{P}(Z ) = Lie_{Z}P$, one finds that
\begin{gather*}
Z \wedge d_{P} Z = 0 \qquad \quad
Z \wedge d_{Q} Z = -2 C_1 \frac{\partial}{\partial{T_1}} \wedge \frac{\partial}{\partial{T_2}} \wedge \frac{\partial}{\partial{T_3}}  .
\end{gather*}
Therefore the 3-vectors $Z \wedge d_{P} Z $ and $Z \wedge d_{Q} Z $  vanish on the submanifold $C_1 = 0 $.
To  interpret this result properly, notice that $Z(C_1) = 2 C_1$. Therefore $Z(C_1)$ is again a Casimir function, and one may  say that the 3-vectors $Z \wedge d_{P} Z $ and $Z \wedge d_{Q} Z $  vanish on the submanifold $Z(C_1) = 0 $. $\square$
\end{paragraph}

\medskip
Let us expand the above example, by introducing the following class of vector fields.
\begin{definition}
Let $M$ be a bihamiltonian manifold of dimension $m = 2q + r + 1 $ endowed with a Poisson pencil $Q- \lambda P$ with $r$ common Casimir functions $( C_1, C_2, \dots, C_r )$, and with a single polynomial Casimir function $g(\lambda)$ of degree $q$. A vector field $Z$ such that:
\begin{itemize}
\item The derivatives $Z(C_{a})$ of the common Casimir functions are still common Casimir functions
\item The 3-vectors $Z \wedge d_{P} Z$  and $ Z \wedge d_{Q} Z$  vanish on the submanifold defined by the equations $Z(C_{a})= 0$
\end{itemize}
will be referred to as a field verifying the Frobenius conditions with respect to the Poisson pencil $Q- \lambda P$ (this terminology is borrowed from the theory of differential forms, where the 1-form $\alpha$ is said to verify the Frobenius condition if $\alpha \wedge d\alpha =0 $). 
\end{definition}
\noindent
For our purposes it is sufficient to deal with the case $q=2$. The following Proposition shows that the vector fields that satisfy the Frobenius conditions are relevant to the theory of separation of variables. 

\begin{proposition} Let $M$ be a bihamiltonian manifold of dimension $m=5+r$, endowed with a Poisson pencil $Q- \lambda P$ having  $r$ common Casimir functions $C_{a}$ and a polynomial Casimir function $g(x)$ of degree $2$. Let $X_{p}$ and $X_{s}$ be the bihamiltonian hierarchy defined by the Poisson pencil (possibly up to an irrelevant common constant multiplicative factor). Assume that  there exists a vector field $Z$ satisfying the Frobenius conditions. Then the
roots of the derivative of $g(x)$ along $Z$, 
\begin{equation}
h(x) = Z(g(x)) ,   \label{h}
\end{equation}
are separating coordinates for  $X_{p}$ and $X_{s}$
on the invariant submanifold $\cal{S}$, defined by the equations $Z(C_{a})=0$. 
\end{proposition}

{\it Proof.}  The value of the 3-vector $Z \wedge d_{P} Z$ on the differentials of the coefficients $K_1, K_2, K_3$ of $g(x)$ is the scalar function
\begin{equation*}
 \sum_{cyclic} Z(K_1) ( Z\{K_2,K_3\}_{P} -\{Z(K_2),K_3\}_{P} -\{K_2,Z(K_3)\}_{P} )  ,
 \end{equation*}
where the sum is over the cyclic permutations of $(1,2,3)$.
Since $K_1$ is a Casimir function of $P$ and $ K_2, K_3$ commute with respect to $P$, many terms in the above expression vanish, and 
one obtains
\begin{equation*}
(Z \wedge d_{P}Z)(dK_1, dK_2, dK_3) = E ( -F_{s} +G_{p} )+ F E_{s} - GE_{p} .
\end{equation*}
Let us restrict this equation to the invariant submanifold $\cal{S}$, and notice that  the restriction of the derivatives of $E, F, G $  coincide with the derivatives of the restriction of the functions, because the vector fields $X_p$ and $X_s$ are tangent to $\cal{S}$. As a result, the functions $E, F, G$ satisfy the first Kowalewski separability condition on $\cal{S}$ provided that $Z$ verifies the Frobenius condition on $\cal{S}$. Under replacement of $P$ by $Q$, the same argument shows that $E, F, G$ satisfy the second Kowalewski separability condition as well. $\square$

It follows that the Poisson pencils which are endowed with a vector field verifying the Frobenius conditions may be qualified as \emph{separable pencils}, because they provide the separating coordinates for their bihamiltonian hierarchies. The following example, which is closely related to the Clebsch system, is quite instructive.

\begin{paragraph}{Example 3.}  The vector field $Z$ of Example 2 verifies the Frobenius conditions with respect to the Poisson pencil $Q-\lambda P$ of Example 1 on the invariant submanifold $\{Z(C_1) = 2 C_1 =0\} $. In this case the polynomial Casimir function is
\begin{equation*}
g(x) =\sum_{\a=1}^{3} (x + j_\b)(x + j_\g) T_{\a}^{2} + (x+j_{\a}) S_{\a}^2 .
\end{equation*}
According to Proposition 3, the coefficients of the quadratic polynomial
\begin{equation*}
h(x) = 2 \sum_{\a=1}^{3} (x + j_\b)(x + j_\g) T_{\a}^{2}  
\end{equation*}
verify the Kowalewski separability conditions on the submanifold $C_1=0$. Thus the roots of the algebraic equation
\begin{equation*}
\sum_{\a=1}^{3}  \frac{T_{\a}^{2}}{x + j_\a} = 0 .
\end{equation*}
are separating coordinates for the Clebsch system on $\{C_1=0\}$. They were first used by H. Weber (\cite{web878}) in 1878. The coordinates of Weber are, therefore, a simple outcome of the Kowalewski separability conditions. $\square$
\end{paragraph}

\section{Separable lifts of Poisson pencils}

Before proceeding to our construction of  separating variables for the Clebsch system, we make a last digression to the theory of Poisson pencils, namely to a process which may be called ''lift of a Poisson pencil''. It will be presented here in a form adapted to our example.

\medskip
Let $M$ be the phase space of the Clebsch system, endowed with the Poisson pencil $Q-\lambda P$, and let $\cal{E}$ be an elliptic curve given in parametric form. The coordinates of points of $\cal E$ are elliptic functions of a parameter $\zeta$ varying on a torus $\mathbb{C}/ L$, $L$ being a period lattice. We do not specify initially neither these functions nor the lattice. The reason is that the choice of the curve $\cal{E}$ is irrelevant to the process of lifting, while it will become crucial later in the search of the separating coordinates. At that point, it will be suggested by the separability conditions.  

To lift the Poisson pencil of the Clebsch system (Example 1) to the extended phase space $\hat{ M }= M \times \cal{E}$, let us choose a scalar function $\hat{C}_2$ on $\hat{M}$, and let $\pi : \hat{M} \rightarrow M$ be the canonical projection. 
Note that there exists a unique pair of bivectors  $\hat{P}$  and $\hat{Q}$  on $\hat{M}$ having the following properties:
\begin{itemize}
\item They can be projected from $\hat{M}$ onto $M$ along $\pi$.
\item Their projections are the bivectors $P$ and $Q$ on $M$.
\item The function $\hat{C}_2$ is a common Casimir function of $\hat{P}$  and $\hat{Q}$.
\end{itemize}
\indent
The simplest way to see these properties is to equip the extended phase space with the fibered coordinates $(S_{\a}, T_{\a}, \zeta )$ adapted to the projection, and to define the bivectors  $\hat{P}$  and $\hat{Q}$ by giving the fundamental Poisson brackets of the coordinate functions on $\hat{M}$. The Poisson brackets of the coordinates $(S_{\a}, T_{\a})$ are the same as on $M$:
\begin{gather*}
\{ S_{\a} , S_{\b} \}_{\hat{P}} =\{ S_\a , S_{\b} \}_{P}  \qquad
\{ S_\a , T_{\b} \}_{\hat{P}} =\{ S_{\a} , T_{\b} \}_{P}  \qquad
\{ T_{\a} ,T_{\b} \}_{\hat{P}} =\{ T_{\a} , T_{\b} \}_{P}  .
\end{gather*}
The Poisson brackets of $(S_{\a}, T_{\a})$ with the new coordinate $\zeta$ are
\begin{align*}
\{ S_{\a} , \zeta \}_{\hat{P}} &= - \left(\sum\limits_{\b=1}^3 \frac{\partial{
\hat{C}_2}}{\partial{S_{\b}}}\{S_{\a},S_{\b}\}_{P}+\frac{\partial{
\hat{C}_2}}{\partial{T_{\b}}}\{S_{\a},T_{\b}\}_{P}\right) \left(\frac{\partial{\hat{C}_2}}{\partial{\zeta}}\right )^{-1}\,  \\
\{ T_{\a} , \zeta \}_{\hat{P}} &= - \left(\sum\limits_{\b=1}^3
\frac{\partial{\hat{C}_2}}{\partial{S_{\b}}}\{T_{\a},S_{\b}\}_{P}+\frac{\partial{\hat{C}_2}}
{\partial{T_{\b}}}\{T_{\a},T_{\b}\}_{P}\right) \left(\frac{\partial{
\hat{C}_2}}{\partial{\zeta}}\right )^{-1}\, .
\end{align*}
These relations come from the condition of $\hat{C}_2$ to be a Casimir function. The Poisson brackets of $\hat{Q}$ are defined similarly. This explicit representation allows to check the above claims, and to recognize that the new pencil  $\hat{Q} -\lambda \hat{P}$ inherits all the properties of the Poisson pencil on $ M$.
In particular,
\begin{itemize}
\item The  pencil $\hat{Q} - \lambda \hat{P}$ is a new Poisson pencil, called the lift of $Q- \lambda P$ from $M$ to $\hat{M}$.
\item It has  two common Casimir functions $\hat{C}_1$ and $\hat{C}_2$, and a quadratic polynomial Casimir function $\hat{g}(\lambda) = \hat{K}_1 \lambda^2+\hat{K}_2\lambda+\hat{K}_3 $.
\item   The coefficients of the polynomial Casimir function $\hat{g}$ are $\pi$-related to the coefficients of the polynomial Casimir function $g$ of Clebsch. Thus the two sets of functions coincide in the fibered coordinates $( S_{\a}, T_{\a}, \zeta ) $. The same claim is true for the Casimir function $\hat{C}_1$: it coincides with the function $C_1$ in such coordinates.The Casimir function $\hat{C}_2$ is the function used to implement the process of lifting.
\item  The new Poisson pencil defines two vector fields $\hat{X}_p$ and $\hat{X}_s$ on $\hat{M}$, which are integrable since $\hat{m}=\hat{n}+2\hat{p}$ .
\item The vector fields $X_p$ and $X_s$ of the Clebsch system are the $\pi$-projections of the fields $\hat{X}_p$ and $\hat{X}_s$.
\end{itemize}

As a result, by lifting the Poisson pencil $Q-\lambda P$ from $M$ to
$\hat{M}$ one obtains an integrable extension of the Clebsch system for any choice of the curve $\cal{E}$ and for any choice of the Casimir function $\hat{C}_2$. This
class of lifts, however, is too ample for our purposes. We are interested in a lifted Poisson pencil which possesses a vector field $\hat{Z}$ satisfying the Frobenius 
conditions on the submanifold $\hat{\cal{S}}$ defined by the equations $\hat{Z}(\hat{C}_1) = 0$ and 
$\hat{Z}(\hat{C}_2) = 0$. This will enable one to implement the Kowalewski separability conditions on $\hat{\cal{S}}$. Furthermore, we want $\hat{\cal{S}}$ to be a six-dimensional covering space of the base manifold $M$. Only in this case one can hope to use the separating
coordinates provided by the Kowalewski separability conditions on $\hat{\cal{S}}$ as separating coordinates 
on the entire phase space $M$. Of course, in this process one must take care of the multi-valuedness 
of the canonical projection $\pi$ restricted to $\hat{\cal{S}}$. This problem will be discussed later on. 

\medskip
A realization of the above program requires appropriate choices of the function $\hat{C}_2$ and of the vector fields $\hat{Z}$. To choose $\hat{Z}$ we consider a class of fields depending uniquely on the parameter $\zeta$ of $\cal{E}$. This allows us to reduce the complicated systems of PDE's,  coming from
the Frobenius conditions, to a more tractable system of ODE's. Hence we set
\begin{equation}\label{Z}
\hat{Z} = \sum\limits_{\a=1}^3 A_{\a}(\zeta) \frac{ \partial}{ \partial{S_{\a}}} + B_{\a}(\zeta) \frac{\partial}{ \partial{T_{\a}}} \, ,
\end{equation}
so that  $\hat{Z}$ does not act on the additional coordinate $\zeta$. To choose $\hat{C}_2$, we recall that the Frobenius conditions require that $Z(\hat{C}_1)$ and $Z(\hat{C}_2)$ be again Casimir functions.  
Our choice is to identify $\hat{C}_2$ with $Z(\hat{C}_1)$  and to set $ Z(\hat{C}_2) = 0$. Accordingly, we set
\begin{gather}\label{C2}
\hat{C}_2 =  \sum\limits_{\a=1}^3 B_{\a}(\zeta) S_{\a} +A_{\a}(\zeta) T_{\a} \ , \qquad   \sum\limits_{\a=1}^3 A_{\a} B_{\a} = 0 .   
\end{gather}
In this way the invariant submanifold $\hat{\cal{S}}$ is a six-dimensional covering space of the base manifold $M$, defined by the single equation $\hat{C}_2=0$. 

As a result, the pencil $\hat{Q} - \lambda \hat{P}$, the vector field $\hat{ Z} $, and the invariant submanifold $\hat{\cal{S}}$ are uniquely defined by  six functions $A_{\a}(\zeta)$ and $B_{\a}(\zeta)$. Our aim is to choose these functions in such a way that $\hat{Z}$ satisfies the remaining Frobenius conditions on $\hat{\cal{S}}$. The existence of these functions is stated in the following proposition proved in Appendix 1.

\begin{proposition} \label{ABZ} Assume that $A_{\a}(\zeta)$ and $ B_{\a}(\zeta) $ verify the four quadratic constraints (\ref{quadrics}) listed in the Introduction.
Then the vector field $\hat{Z}$, defined by \eqref{Z}, satisfies the Frobenius conditions with respect to the Poisson pencil  $\hat{Q} -\lambda \hat{P}$ on the invariant submanifold $\hat{\cal{S}}$.
\end{proposition}

This result is the keystone of our construction of separating coordinates for the Clebsch system. It will be fully exploited in the next section. Before, we use Proposition \ref{ABZ}  to specify the elliptic curve $\cal{E}$, the invariant submanifold $\hat{\cal{S}}$, and the canonical projection $\pi: \hat{\cal{S}} \rightarrow M$. 

First, note that the functions $A_{\a}(\zeta), B_{\a}(\zeta)$, verifying the constraints (\ref{quadrics}), are rational  
on a spatial curve in $\mathbb{C}^3(v_1, v_2, v_3)$ given by the equations 
\begin{gather} \label{E}
v_{1}^{2} - v_{2}^{2} =  j_1 - j_2\, , \quad
v_{1}^{2} - v_{3}^{2} =  j_1 - j_3 \, .
\end{gather}
Since $j_1,j_2, j_3$ are distinct, the above curve is elliptic. Then one can write 
\begin{equation}\label{param}
A_{\a} = c_{\a} v_{\b} v_{\g}\, , \quad  B_{a} = c_{\a} v_{\a} ,
\end{equation}
assuming that the coefficients $c_{\a}$ satisfy the constraints
$\sum\limits_{\a=1}^3 c_{\a}^{2} = 0 \,, \quad \sum\limits_{\a=1}^3 j_{\a} c_{\a}^{2} = 0$. So 
\begin{equation*}
c_{\a}^{2} = \frac{\rho^2}{( j_{\a} - j_{\b})( j_{\a} - j_{\g})} ,
\end{equation*}
where $\rho$ is an arbitrary nonzero factor. Since all the equations of the present theory are gauge-invariant with respect to the choice of the gauge function $\rho$, one may assume, without loss of generality, that $\rho$ is a constant.
In the sequel we will choose $\rho=1$. The conclusion is that one can identify $\cal{E}$ with the above elliptic curve, proving the claim that $\cal{E}$ is fixed by the separability conditions. 

Next, the submanifold $\hat{\cal{S}}\subset {\hat{M}}$ is fully specified as the zero-locus of the Casimir function $\hat{C}_2$. 
By inserting the above parametric representations of $ A_{\a}$ and $ B_{\a}$ into the first equation in \eqref{C2}, one obtains
\begin{equation}\label{vinc}
\sum\limits_{\a=1}^3 c_{\a} (v_{\b} v_{\g} T_{\a} + v_{\a}S_{\a}) =0 .
\end{equation}
Jointly with \eqref{param}, \eqref{E}, this equation defines $\hat{\cal{S}}$.  Finally, the restriction of the projection $\pi$ onto the submanifold $\hat{\cal{S}}$ have the following property, which follows from Proposition 4. 

\begin{corollary}  The invariant submanifold $\hat{\cal{S}}$ is a 8-fold covering of the base manifold $M$. 
\end{corollary}
{\it Proof.}  The curve $\cal{E}$ can be transformed to the canonical Weierstrass form in ${\mathbb C}^2(Z,Y)$ 
\begin{equation} \label{hatE}
\hat{E}: \qquad Y^{2} =4 ( Z+j_1)(Z+j_2 )( Z+ j_3) 
\end{equation}
by means of the birational transformation
\begin{equation} \label{v_alpha_ZY}
v_{\a} = \frac{ (Z+j_{\a})( Z+j_{\b}) - (Z+j_{\b})( Z+j_{\g}) +( Z+j_{\g})( Z+j_{\a})}{Y} . 
\end{equation}
Indeed, under \eqref{hatE}, substituting \eqref{v_alpha_ZY} into \eqref{E} converts the latter into identities.

Next, substituting \eqref{v_alpha_ZY} into \eqref{vinc} yields 
\begin{equation*}
P_{2}(Z) Y +P_{4}(Z) = 0 ,
\end{equation*}
where  $P_{2}(Z)$ and $P_{4}(Z)$ are certain polynomials of degree 2 and 4 respectively, with coefficients depending 
linearly on $(S_{\a}, T_{\a})$. (These polynomials will be presented  in explicit form in the second part of the paper.)
Eliminating from here $Y$ by using \eqref{hatE} gives a polynomial equation of degree eight in $Z$. 
Hence, to any given point of $M$ there correspond eight points on $\hat {\cal S}$. $\square$

\bigskip
From a different perspective, this corollary justifies the claim, made in the Introduction, that the curve in $\mathbb{P}^5$ defined 
by the four quadrics (\ref{quadrics}) is cut by the plane  (\ref{plane}) in eight points. Indeed, the previous argument shows that
there is a one-to-one correspondence between the intersections in $\mathbb{P}^5$ of the quadrics (\ref{quadrics}) with the plane  (\ref{plane}) 
and the inverse images on $\hat{\cal{S}}$ of a point  in $M$. From this point of view it is interesting to interpret, within the geometry  of 
the lift of the Poisson pencil, the meaning of the parameter $v$ associated by (\ref{v}) to any intersection point in $\mathbb{P}^5$.
A simple substitution of the parametric equations of $ A_{\a}$ and $ B_{\a}$ into (\ref{v}) shows that $v$ is the function
\begin{equation}\label{v2}
v= v_{\a}^2 - j_{\a},
\end{equation}
associated to anyone of the eight inverse images of $(S_{\a}, T_{\a})$ on $\hat{\cal{S}}$. This remark gives us the chance to summarize
the part of the study concerning the lifting process and the Kowalewski separability conditions in $\hat{M}$ as follows. Each point $(S_{\a}, T_{\a})$ 
of the phase space $M$ of Clebsch define eight points on the submanifold $\hat{\cal{S}}$, and therefore eight values of the function $v$. 
By the Kowalewski separability conditions the function $v$ is one of the separating coordinates
of the extensions $\hat{X}_{p}$ and $\hat{X}_{s}$ of the Clebsch system (as we shall see in the next section). Therefore, each solutions 
of the Clebsch system on $M$ selects eight solutions of the separation equations for $\hat{X}_{p}$ and $\hat{X}_{s}$ on $\hat{\cal{S}}$.
The interesting property is that this correspondence may be inverted: from each suitable set of eight solutions of the separation equations of $\hat{X}_{p}$ and $\hat{X}_{s}$ on $\hat{\cal{S}}$ one can reconstruct a solution of the Clebsch equations of motion. This delicate mechanism will be studied in the
second part of the paper, where it will be shown that one can exploit such a 8:1 correspondence to write the solutions of the Clebsch equations of motion
in terms of theta-functions. Before addressing this reconstruction problem, however, one has still to identify the separating coordinates and the form of the equations of the vector fields $\hat{X}_{p}$ and $\hat{X}_{s}$ in these coordinates, at least on $\hat{\cal{S}}$. This is the task of the last section of this first part.

\section{Separation of variables in action}
As we know, on the extended space $\hat{M} = M \times \cal{E}$  the vector field $\hat{X}_p$ is characterized by two main properties:
\begin{itemize}
\item  Its projection onto $M$ is the vector field $X_p$ of the Clebsch system.
\item The function $\hat{C}_2$ is one of its integral of motion. 
\end{itemize} 
We also know that $\hat{X}_p$ is an integrable bihamiltonian vector field, and that its restriction to the invariant submanifold $\hat{\cal{S}}= \{\hat{C}_2=0\}$ is separable. This means that in the coordinate system formed by the polynomial Casimir functions and by 
the roots $x_1, x_2$ of the derivative (\ref{h}) of the polynomial Casimir function, the restriction of $\hat{X}_p$ to $\hat{\cal{S}}$ admits the expansion (coming 
from Eq. \ref{Stack})
\begin{equation}
( x_1 - x_2 ) \hat{X}_p = \psi_1 \frac{\partial}{\partial{x_1}} - \psi_2 \frac{\partial}{\partial{x_2}} , \label{expan}
\end{equation}
where the component $\psi_1$ does not depend on $x_2$, and $\psi_2$ does not depend on $x_1$.  
In this section we compute these components explicitly and, using them, bring the differential equations defined 
by the fields $\hat{X}_p, \hat{X}_s$ to quadratures.

\paragraph{A change of coordinates.} To reach this goal it is convenient to use the following redundant coordinate system  
on $\hat{M}$. We will keep using the coordinates $v_1,v_2,v_3$ parameterizing the points of the fiber $\cal{E}$ and introduce six new coordinates 
\begin{gather} \label{XY}
X_{\a} = B_{\a} S_{\a} = c_{\a} v_{\a} S_{\a}, \quad 
Y_{\a} = A_{\a}T_{\a}  = c_{\a} v_{\b} v_{\g} T_{\a} .
\end{gather}
Then the equation of $\hat{\cal{S}}\subset\hat{M}$ and of the field $\hat{Z}$ take the following simple forms
\begin{equation} \label{vinc1}
\hat{\cal{S}} : \quad \sum_{\a=1}^{3} (X_{\a} + Y_{\a}) = 0\, ,  \quad
\hat{Z}=\sum_{\a=1}^{3} c_{\a}^2 \left( \frac{\partial}{\partial{X_{\a}}} + \frac{\partial}{\partial{Y_{\a}}} \right) ,
\end{equation}
whereas the expressions of the polynomial Casimir functions become
\begin{gather*}
\hat{C}_1 = \frac{1}{v_1v_2 v_3} \sum_{\a=1}^{3} d_{\a} X_{\a} Y_{\a} \, ,\quad
\hat{C}_2= \sum_{\a=1}^{3} (X_{\a} + Y_{\a}) \, , \\
\hat{g}(x) = \sum_{\a=1}^{3} d_{\a} \frac{(x+j_{\b})(x+j_{\g})}{(v+j_{\b})(v+j_{\g})} Y_{\a}^2 + \sum_{\a=1}^{3} d_{\a} \frac{(x+j_{\a})}{(v+j_{\a})} X_{\a}^2 \,, \qquad d_{\a} := \frac{1}{c_{\a}^2} .
\end{gather*}
The function $v$ has been defined in (\ref{v2}). As shown before, it coincides with the function $v$ defined by formula (\ref{v}) in Introduction.

\paragraph{The vector field $\hat{X}_p$.}  There are infinitely many vector fields on $\hat{M}$ whose projection onto $M$ give $X_{p}$, 
they depend on an arbitrary function. The whole class of these fields is specified by the equations  
\begin{align*}
\dot{v}_{\a } &= \frac{1}{v_{\a}} \xi \, , \\
\dot{X}_{\a } &= \frac{1}{v_{\a}^2} X_{\a} \xi + \frac{i }{v_1 v_2 v_3} (j_{\g} -j_{\b})^{2} Y_{\b} Y_{\g}\, , \\
\dot{Y}_{\a } &= \frac{v_{\b}^2 +v_{\g}^2}{v_{\b}^2 v_{\g}^2} Y_{\a} \xi + \frac{i }{v_1 v_2 v_3} (j_{\g} - j_{\b}) (v_{\g}^2 X_{\b} Y_{\g}  - v_{\b}^2 X_{\g} Y_{\b} ) , 
\end{align*}
where $\xi$ is an arbitrary scalar function on $\hat{M}$. The first equation is due to the equations defining the curve $\cal{E}$. The second equation is obtained in three steps. First, one substitutes 
$\dot{S}_{\a}$ by the corresponding differential equations of the field $X_p$ into the identity $\dot{X}_{\a } = \dot{B}_{\a} S_{\a}+ B_{\a} \dot{S}_{\a} $; then, one replaces $(S_{\a}, T_{\a} )$ by the new coordinates  $(X_{\a}, Y_{\a} )$; finally, one exploits the definitions of the functions $ ( A_{\a}, B_{\a} )$ and of the parameters $ c_{\a} $ given in the previous section. The third equation is obtained in a similar way. Among the vector fields of this class, $\hat{X}_p$ is characterized by the property of leaving the function $\hat{C}_2$ invariant. This condition yields  
$ \xi =  i\, v_1 v_2 v_3\, U /V $,
where
\begin{align}
\begin{split}
U &=  \sum\limits_{\a=1}^3  (j_{\b} - j\g) (v_{\g}^2 X_{\b} Y_{\g} - v_{\b}^2 X_{\g} Y_{\b} - (j_{\b} - j\g) Y_{\b} Y_{\g} ) \\  \label{UV}
V &= \sum\limits_{\a=1}^3 v_{\b}^2 v_{\g}^2 X_{\a} + v_{\a}^2 (v_{\b}^2 + v_{\g}^2) Y_{\a} \,. 
\end{split}
\end{align}
Restrictions of $U,V$ onto the submanifold $\hat{\cal{S}}$ describe the restriction of the field $\hat{X}_p$. In this connection, one observes that the restriction of 
$U$ admits the following factorization 
\begin{equation}
U \big |_{\hat{\cal{S}} } = L M, \qquad
L = \sum\limits_{\a=1}^3 (x_1+j_{\a}) Y_{\a}\, , \quad  M = \sum\limits_{\a=1}^3  j_{\a} ( X_{\a} + Y_{\a} ) \, \,; \label{LM}
\end{equation}
hence the function $\xi$ simplifies to
$\xi =   i \, v_1 v_2 v_3 M L/V$, 
and the components of $\hat{X}_p$ along the curve $\cal{E}$ become
\begin{equation} \label{vp}
\hat{X}_{p}(v_{\a}) = i  v_{\b} v_{\g} M \frac{L }{V} \, . 
\end{equation}
These expressions play a key role in the process of separation of variables.

\paragraph{Separating coordinates.} In the new coordinate system 
$(v_\alpha, X_\alpha, Y_\alpha)$, $\alpha=1,2,3$, the equation \eqref{SP}, whose roots give the separating coordinates $x_1,x_2$  according to Proposition 3, takes the following symmetric form
\begin{equation}
\sum\limits_{\a=1}^3 \frac{(x+j_{\a})}{(v+j_{\a})} X_{\a}  + \frac{(x+j_{\b})(x+j_{\g})}{(v+j_{\b})(v+j_{\g})} Y_{\a} = 0 .  \label{PS2}
\end{equation}
One of its roots, say $x_{1}$, is the function $v $ introduced above,  and $x_{2}$ is related to $x_{1}$ by a Mobius transformation, namely  
\begin{equation}  \label{Mo}
x_{1}  =  v , \quad x_{2} = \frac{a x_1 + b}{c x_1 + d} , 
\end{equation}
where
\begin{gather*}
a = \sum_{\a=1}^{3} j_{\a} X_{\a} , \quad    b = -\sum_{\a=1}^{3} j_{\b}j_{\g} X_{\a} + j_{\a} ( j_{\b} + j_{\g} ) Y_{\a} ,     \quad  c = \sum_{\a=1}^{3} Y_{\a} ,  \qquad  d = \sum_{\a=1}^{3} j_{\a} Y_{\a} .
\end{gather*}
The first relation in \eqref{Mo} holds because setting $x=v$ reduces the equation \eqref{PS2} to the constraint 
\eqref{vinc1}. The second relation follows from the Vi\'ete formulas. 
Note that on $\hat{\cal{S}}$ the determinant of the Mobius transformation
\begin{equation*}
a d - b c = \sum\limits_{\a=1}^3 (j_{\a}-j_{\b})(j_{\a}-j_{\g})X_{\a}Y_{\a} 
\end{equation*}
is proportional to the first Casimir function $\hat{C}_1$. Thus this transformation is regular on $\hat{C}_2=0$ outside of the singular variety $\hat{C}_1=0$. Next, the determinant, which is a quadratic form of the coordinates, on $\hat{\cal{S}}$ factorizes into the product of two linear factors:   
\begin{equation}
a d - b c = L N , \qquad N= \sum\limits_{\a=1}^3 (x_2+j_{\a}) X_{\a} \qquad \mbox{ on \; $\hat{\cal{S}}$ }. \label{N}
\end{equation}
Other relevant property is that on $\hat{\cal{S}}$ the difference $ x_1-x_2 $  is the rational function
\begin{equation} \label{diff}
x_1- x_2  = \frac{V }{ L} ,
\end{equation} 
the linear forms $L$ and $V$ have been defined in \eqref{LM}, \eqref{UV}. These observations are sufficient to reduce the Clebsch system to quadratures.

\paragraph{The components $\psi_1$ and $\psi_2$.} According to  \eqref{expan}, the components $\psi_1, \psi_2$ of the field 
$\hat{X}_p$ can be written as 
$\psi_1= (x_1-x_2)x_{1 p}$, $\psi_2= (x_2-x_1)x_{2 p}$.
Thus, to compute them one has to find first 
$x_{ 1 p}, x_{ 2p}$. The former is easy to find: 
\begin{equation*}
x_{ 1 p} = v_{p} = 2 v_{\a} v_{\a p} = 2 i \, v_1 v_2 v_3 M\, \frac{L}{R} .
\end{equation*}
Due to \eqref{diff}, the above relation immediately yields
$\psi_1 = 2 i \, v_1 v_2 v_3 M$.

In this way we obtain $\psi_1$ as a function of the redundant coordinates $( X_{\a}, Y_{\a}, v_{\a} )$. To express it in terms of the separating coordinates, that is, as a function 
of the polynomial Casimir functions and of $ x_1$, we use the following (easily checked) identity
\begin{gather*}
M^2= \hat{g}(x_1)+ 2\, v_1v_2v_3 \hat{C}_1 
\end{gather*}
and the identity $v_1v_2v_3=\sqrt{(x_1+j_1)(x_1+j_2)(x_1+j_3)}$.

Determination of $x_{2p}$ is longer and computationally more challenging.
To compute it one might derive the Mobius 
transformation  \eqref{Mo} along $\hat{X}_{p}$, then insert  the derivatives of $x_{1}, X_{\a }, Y_{\a }$ into the resulting expression, and
take into account the constraint \eqref{vinc1}.
Next, one needs to replace $( X_{\a }, Y_{\a } )$ by their expressions in terms of the separating coordinates. In this form the computation cannot be brought to an end due to overwhelming complexity of the resulting expressions (containing thousands of terms). 

The way to bypass this difficulty is suggested by the study of the polynomial Casimir functions of the Poisson pencil. As we have seen, on $\hat{\cal{S}}$
\begin{gather*}
\sqrt{\Phi(x_1)} \,\, \hat{C}_{1} = L N , \qquad       \hat{g}(x_1) + 2 \sqrt{\Phi(x_1)} \, \hat{C}_{1} = M^2  , \qquad  \Phi(x) := (x+j_1)(x+j_2)(x+j_3). 
\end{gather*}
These factorization properties 
suggest introducing a new coordinate system on $\hat{\cal{S}}$ including the separation coordinates $x_1, x_2$ and the linear forms $L, M, N$. The system is completed by choosing, as a sixth coordinate, $X\,: =\sum (x_1+j_{\b})(x_1+j_{\g}) X_{\a}$. 


The transformation $(X_{\a}, Y_{\a}) \to (x_1, x_2, L, M, N, X)$ is given by the five equations above defining $x_2, L, M, N, X$, and by the constraint equation \eqref{vinc1}. Solving them with respect to $X_{\a}, Y_{\a}$, one obtains the expressions of the inverse transformation
\begin{align}\label{XYXY}
\begin{split}
d_{\a} X_{\a} &= \frac{(x_1+j_{\a})(x_1+x_2+j_{\b}+j_{\g})}{x_1-x_2} (L-M+N) + (x_1+j_{\a}) N +X \, , \\
d_{\a} Y_{\a} &= \frac{(x_1+j_{\b})(x_1+j_{\g})}{x_1-x_2} (L-M+N) + (x_2+j_{\a}) L +X  \, .
\end{split}
\end{align}
The use of the new coordinates drastically simplifies computation of $x_{2p}$, which is based on the equation
\begin{equation*}
x_{2p} = x_{1p} - \frac{L V_{p} - V L_{p}}{L^2} = x_{1p} +  (x_1-x_2) \frac{L_{p}}{L} - \frac{V_{p}}{L} .
\end{equation*}
The derivatives $L_{p}, V_{p}$ of the linear forms $L, V$ along $\hat{X}_{p}$ are still rather complicated, but after their insertions into the above equation an impressive cancellation of terms takes place, which leads to the following simple relation 
\begin{equation*}
(x_2-x_1)x_{2 p} =i \, \left( \Phi(x_2) \frac{L}{\sqrt{\Phi(x_1)}} - \frac{\sqrt{\Phi(x_1)}}{L}  N^2  \right).   
\end{equation*}
Therefore
\begin{gather*}
\psi_2 = i \left(\Phi(x_2) \frac{L}{\sqrt{\Phi(x_1)}} -   \frac{\sqrt{\Phi(x_1)}}{L}  N^2 \right) .
\end{gather*}
To complete the computation of $x_{2p}$, it remains to express  $L,N$ in terms of the polynomial Casimir functions. This is performed by using the third noticeable relation
\begin{equation*}
\hat{g}(x_2) = \frac{ \Phi(x_2)}{\Phi(x_1)} L^2 + N^2  ,
\end{equation*}
which is obtained in the same way as $x_{2 p}$ was.

\paragraph{Reduction to quadratures.}  The above expressions for $\psi_1$ and $\psi_2$ may be recast in a more symmetric form by rescaling them by the factor $\hat{C}_1$. Indeed, a simple algebraic manipulation yields
\begin{equation*}
\hat{C}_1 \psi_1 = i \, M (  M^2 - \hat{g}(x_1) ),  \qquad
\hat{C}_1 \psi_2 = i\,  N ( \hat{g}(x_2) - 2  N^2  ) . 
\end{equation*}
Setting here $w=M$ and $W =- \sqrt{2} N$, we arrive at the final form: 
\begin{equation*}
\hat{C}_1 \psi_1 = i w ( w^2 - \hat{g}(x_1) ), \qquad
\hat{C}_1 \psi_2 = i \frac{W}{\sqrt{2}} (W^2 -\hat{g}(x_2) ) . 
\end{equation*}
As a consequence of the above expressions, the new variables $w, W$ are manifestly related to the constants of motion and the separating coordinates as described in equations (\ref{curves}). The latter define the pair of algebraic curves $C$ and $K$ associated with the Clebsch system by the theory of Kowalewski separability conditions. 

Finally, to bring the Clebsch system to quadratures we plug the above representation of $\psi_1$ and $\psi_2$ into the Abelian form of the equations of motion (\ref{Ab2}) provided by the theory of Kowalewski separability conditions. The result is given by the equations (\ref{Ab}) in Introduction. Their integration leads to the quadrature formula (\ref{A-P_0}). A description of its properties and its inversion will be given in the second part of the paper.  

\paragraph{Reconstruction formulae.} We now express the original variables $S_\a, T_\a$ on the phase space $M$ in terms of the separating variables $(x_1,w), (x_2,W)$. 

\medskip
Let us recall that the  Poisson tensor $P$, described in Example 1, has two Casimir functions $C_1$ and $K_1$, and that its symplectic leaves are, accordingly, four-dimensional submanifolds of $M$. Consider the generic symplectic leaf $\{C_1=e\ne 0, K_1=f\}$ and notice that the inverse image of it with respect to the projection $\pi :\hat{\cal{S}} \rightarrow M$ is the four-dimensional submanifold of $\hat{\cal{S}}$ satisfying the conditions  $\hat{C}_1=e$, $\hat{K}_1=f$. In the new coordinates $(x_1,x_2,L,M,N,X)$ on $\hat{\cal{S}}$ this submanifold is defined by equations
\begin{equation} \label{ef}
e = \frac{L N}{\sqrt{\Phi(x_1)}}\,,  \qquad
f =\frac{L}{\Phi(x_1)} \left( L (2x_2+x_1+ j_1+j_2+j_3) + 2 X \right) + \frac{(L-M+N)^2}{(x_1-x_2)^2}\, .
\end{equation}
Solving this system with respect to $L,X$, one obtains them as functions of $x_1,x_2,M,N,e,f$. 

On the other hand, recall that the projection $\pi :\hat{\cal{S}} \rightarrow M$ is described by equations \eqref{XYXY}, which, in view of \eqref{XY}, can be written in the form
\begin{align} \label{recon}
\begin{split}
S_{\a} &= c_{\a}\left(  \frac{\sqrt{x_1+j_{\a}} (x_1+x_2+j_{\b}+j_{\g} )}{x_1-x_2} (L-M+N) + \sqrt{x_1+j_{\a}} \, N +\frac{X}{\sqrt{x_1+j_{\a}}}\right), \\
T_{\a} &= c_{\a}\left(  \frac{\sqrt{(x_1+j_{\b})(x_1+j_{\g})} }{x_1-x_2} (L-M+N) + \frac{\sqrt{x_2+j_{\a}}}{ \sqrt{(x_1+j_{\b})(x_1+j_{\g})} }\, L +\frac{X}{\sqrt{(x_1+j_{\b})(x_1+j_{\g})}} \right) .
\end{split}
\end{align}
Replacing here $L,X$ by the solutions $L(x_1,x_2,M,N,e,f)$, 
$X(x_1,x_2,M,N,e,f)$ of \eqref{ef}, then setting, as above, $M=w$, $N=-W/\sqrt{2}$, we arrive at {\it algebraic} expressions for $S_\a, T_\a$ in terms of $(x_1,w), (x_2,W)$, on the symplectic leaf 
$\{C_1=e, K_1=f\}$. (Their explicit form is not too long, but we prefer not to present it here.) Combined with solutions (inversion) of the quadratures \eqref{Ab}, these expressions give us solutions of the Clebsch system.   

\paragraph{Remark.} The reconstruction formulae \eqref{recon} give  $S_\a, T_\a$ as rational functions of $x_1, \sqrt{\Phi(x_1)}, w$, which are meromorphic on the curve $C$ in \eqref{curves}, and of $x_2,W$, which are meromorphic on $K$. 
However, these formulae involve also the radicals 
$\sqrt{x_1+j_{\a}}$, $\a=1,2,3$, which are not meromorphic on $C$, but on its 4-fold unramified covering ${\cal D}\to C$, 
$\cal D$ being a genus 9 curve whose properties are described in detail in Part II of the paper. As a result, for a generic pair of points $(x_1, w) \in C$, $(x_2, W) \in K$, the formulae \eqref{recon} yield not one but 4 sets $(S_\a, T_\a)$, which, however, are only different by signs.   
    
Indeed, taking squares of the right hand sides of \eqref{recon}, we obtain rational functions of $x_1, \sqrt{\Phi(x_1)}, w$ and $x_2,W$. Thus, $S_\a^2, T_\a^2$ are determined by a pair of points $(x_1, w) \in C$, $(x_2, W) \in K$ uniquely.     

Finally, observe that, by their construction, expressions \eqref{recon} give the same values of $S_\a^2, T_\a^2$ for each pair $(x_1, w), (x_2, W)$ of the eight sets of separating variables obtained above.    
 
\paragraph{Conjugate momenta.} As was shown in a preliminary version of this paper \cite{MaSk},  the separation coordinates $x_1$ and $x_2$ commute with respect to both Poisson brackets on $\hat{M}$ defined by the Poisson tensors $\hat{P}$ and $\hat{Q}$. This noticeable property brings our separation procedure into the framework of the standard Hamilton--Jacobi theory. (Curiously, after more than a century it is still unknown whether the same commutativity property is shared by the separating variables constructed by K\"otter in \cite{kot892}). Without outlining here the transition from the algebraic geometric picture developed so far to the Hamilton-Jacobi picture of Classical Mechanics, we simply recall that for $x_1,x_2$, the conjugate momenta with respect to the Poisson tensor $\hat{P}$ are
\begin{gather*}
y_1 = \frac{M}{2 \sqrt{\Phi(x_1)}}\, ,   \quad y_2 = \frac{L}{\sqrt{\Phi(x_1)}}.
\end{gather*}
So the algebraic coordinates $w$ and $W$ are related to the conjugate momenta according to $w=2 \sqrt{\Phi(x_1)} \; y_1$ and $W=- \frac{ \hat{C}_1}{\sqrt{2}y_2}$.  The equations of the curves $C$ and $K$ become then the equations giving the Hamiltonians $H_p$ and $H_s$  as functions of the separating coordinates $x_1$, $x_2$, and of their conjugate momenta $y_1$,$y_2$. From these equations one may fully reconstruct the description of the Clebsch system from the viewpoint of the classical Hamilton-Jacobi theory.

\section{Final remarks}
We believe that the above exposition brings two new ideas in the field of integrable systems.

The first one is the concept of separable Poisson pencils. They provide both an integrable Hamiltonian system
and the coordinates which allow to solve it by separation of variables. Such pencils are characterized by two properties: 
\begin{itemize}
\item Their polynomial Casimir functions satisfy the integrability condition $m=n+2p$;
\item In the phase space there exist vector fields $Z_{a}$ (their number equals the number of polynomial Casimir functions) which solve a set of 
cohomological conditions of the type $Z_{a}\wedge d_{P}Z_{a}=0$ ( and similar, but more complicated, conditions with respect to the second Poisson tensor $Q$). 
\end{itemize}
This paper gives just a first example of this occurrence. For the Clebsch system the pencil has a single
polynomial Casimir function of degree two. Nevertheless, the tools used in the paper allow extensions to pencils with an arbitrary number of polynomial Casimir functions. Naturally, in this case the cohomological conditions increase in number and become more complicated. 
For instance, if the pencil has two polynomial Casimir functions of degrees 1 and 2, there are 12 such conditions.  
However, the idea is always the same: the cohomological conditions imply the Kowalewski separability conditions, and therefore guarantee 
the existence of a set of separating coordinates for the integrable bihamiltonian systems defined by the polynomial Casimir functions. 

The second idea is the lifting. Several examples show that in order to be able to solve the previous cohomological conditions one must accept 
the idea of constraining the dynamical vector fields to suitable invariant submanifolds. From our standpoint, this is the correct interpretation of the 
pioneering work of Weber \cite{web878}. The lift counterbalances the need for a constraint. The process of lifting is a topic to be still 
thoroughly investigated, yet a point seems to emerge clearly from the few examples we have studied so far. There is a class of integrable and separable 
bihamiltonian systems for which there are two distinct algebraic curves supporting the separating coordinates. In other words, the Clebsch system is not an isolated example. For this reason we believe that the study of the problem of inversion of the quadratures involving two different algebraic curves, which is considered in the second part of the paper, deserves 
special attention.

\subsection*{Acknowledgments} The authors acknowledge the support and hospitality of Department of Mathematics of Universit\'a di Milano Bicocca, where a part of this work has been made. The contribution of Y.F was partially supported by the Spanish MINECO-FEDER Grant MTM2015-65715-P and the Catalan grant 2017SGR1049.  T. Skrypnyk is grateful to the late B. Dubrovin for support and discussions.

\section*{Appendix : Proof of Proposition 4}
On the seven-dimensional manifold $\hat{M}$, the 3-vector  $\hat{Z} \wedge d_{\hat{P}}\hat{ Z} $  
has 35 components. They are computed by inserting any triple of coordinate functions $(J,K,L)$ into the equation
\begin{equation*}
(Z \wedge d_{\hat{P}}Z)(dJ, dK, dL) = \sum_{cyclic}Z(J) \left( Z\{K,L\}_{P} -\{Z(K),L\}_{P}
-\{K,Z(L)\}_{P} \right) ,
\end{equation*}
where the sum is over the cyclic permutations of the coordinate functions.  Fifteen components vanish due to the assumption that $\hat{C}_2$ is a Casimir function and that $\hat{Z}(\hat{C}_2) =0$ ( just set $L=\hat{C}_2$ in the previous equation). Hence, the problem is to study the vanishing of the remaining 20 components, both in the case of the Poisson tensor $\hat{P}$ and in the case of the Poisson tensor $\hat{Q}$. Since
the unknown functions $A_{\a}(\zeta)$ and $B_{\a}(\zeta)$ do not depend on the coordinates $(S_{\a}, T_{\a})$, each of the above 40 equations split, by separation of variables, into 3 ODE's. So, one has to study the solvability of a system of 120 ODE's in 6 unknown functions. This study is split in four Lemmas.
For brevity, we shall agree to denote by a prime the derivative with respect to the coordinate $\zeta$ of any function defined on the curve $E$. 

\begin{lemma}\label{I} If the functions  $ A_{\a}(\zeta)$ and $ B_{\a}(\zeta) $ verify the algebraic constraints
\begin{equation*}
\sum\limits_{\a=1}^3 A_{\a} B_{\a} = 0, \qquad  \qquad
\sum\limits_{\a=1}^3 B_{\a}^{2} = 0 ,
\end{equation*}
and the differential equations
\begin{equation*}
B_{\b} \sum_{\a=1}^{3}A_{\a}A_{\a}^{'} - B_{\b}^{'} \sum_{\a=1}^3
A_{\a}^{2}  - A_{\b} \sum_{\a=1}^3 B_{\a} A_{\a}^{'}= 0 ,
\end{equation*}
the 3-vector $\hat{Z} \wedge d_{\hat{P}}\hat{ Z} $ vanishes on  $\hat{\cal{S}}$.\end{lemma}
{\it Proof.} Let us choose  $J=T_1, K=T_2, L=T_3$. By taking into account the constraint $\hat{C}_2=0$, one readily finds 
that this component vanishes if and only if the functions $B_{\a}(\zeta)$ satisfy the differential equations
\begin{equation*}
B_{\b} \sum_{\a=1}^3 B_{\a}B_{\a}^{'} - B_{\b}^{'} \sum_{\a=1}^3 B_{\a}^{2}=0 .
\end{equation*}
In the same way, one finds that the component $J=S_1, K=S_2, L=S_3$ vanishes if and only if the differential equations written in the Lemma are satisfied.
To control the remaining components, let us preliminarily notice that the algebraic constraints listed in the Lemma imply the existence of a 
multiplier $\rho$ such that
\begin{equation*}
\rho B_{\a} = B_{\b} A_{\g} - B_{\g} A_{\b} .
\end{equation*}
This multiplier is  one of  the roots of the quadratic equation $\rho ^{2} + \sum_{\a} A_{\a}^{2}=0 $ since, otherwise, the above
linear system would have only the trivial solution $B_{\a} = 0$. By plugging the
last relation into the  differential equations one readily put them in the form
\begin{equation*}
\rho B_{\a}^{'} = B_{\b} A_{\g}^{'} - B_{\g} A_{\b}^{'} .
\end{equation*}
They allow to check that the remaining 18 components of the 3-vector $\hat{Z} \wedge d_{\hat{P}}\hat{ Z}$ 
vanish on the surface $\hat{C}_2=0$, without any further condition. $\square$

\begin{lemma} \label{II} Assume that the seven functions  $ A_\a(\zeta)$, $ B_\a(\zeta) $ and $v(\zeta)$  satisfy the ten algebraic constraints
\begin{align*}
A_{\a}^{2} + (v+j_{\b}) B_{\g}^{2} + (v+j_{\g}) B_{\b}^{2} &= 0, \\
(v+j_{\a}) A_{\a} B_{\b} - (v+j_{\b}) A_{\b} B_{\a}  &=0, \\
A_{\a} A_{\b} - (v+j_{\g}) B_{\a} B_{\b} &=0,  \\
\sum\limits_{\a=1}^3 (v+j_{\a}) A_{\a}^{2} &=0 .
\end{align*}
Then the  3-vector $\hat{Z} \wedge d_{(\hat{Q}+v(\zeta) \hat{P})}\hat{ Z}$ vanishes on $\hat{\cal{S}}$.
\end{lemma}
{\it Proof.} One has to repeat the computation done before by replacing the Poisson tensor $P$ with the Poisson pencil $\hat{Q}+v \hat{P}$. 
One finds that the component relative to the choice  $J=T_1, K=T_2, L=T_3$ vanishes,  on the surface $\hat{C}_2=0$, 
if and only if the following system of differential equations
\begin{gather*}
\begin{split}
B_{\a}^{'} ( A_{\a}^{2} + (v+j_{\b}) B_{\g}^2 + (v+j_{\g}) B_{\b} ^2) +\\ + ( A_{\a} A_{\b} - (v+j_{\g}) B_{\a} B_{\b} ) B_{\b}^{'} + ( A_{\a} A_{\g} - (v+j_{\g}) B_{\a} B_{\g} ) B_{\g}^{'} = 0 
\end{split}
\end{gather*}
is satisfied. Similarly, one recognizes that the choice $J=S_1, K=S_2, L=S_3$ leads to the differential equations 
\begin{gather*}
\begin{split}
B_{\a}^{'} \sum_{\a=1}^3 (v+j_{\a}) A_{\a}^{2}+ ((v+j_{\a}) A_{\a} B_{\b}- (v+j_{\b}) A_{\b} B_{\a} ) A_{\b}^{'} +\\ + ( (v+j_{\a}) A_{\a} B_{\g}  - (v+j_{\g}) A_{\g} B_{\a} ) A_{\g}^{'}  = 0 .
\end{split}
\end{gather*}
One may notice that the coefficients of these  differential equations are the quadratic polynomials listed in Lemma \ref{II}. So the algebraic constraints imposed by the Lemma define a singular solution of these differential equations.The inspection of the remaining 18 components confirms this occurrence in general. $\square$

\medskip
Since the vector field $\hat{Z}$ does not affect the coordinate $\zeta$, to say that the 3-vectors  $\hat{Z} \wedge d_{\hat{P}}\hat{ Z} $  and $ \hat{Z} \wedge d_{(\hat{Q}+v(\zeta) \hat{P})}\hat{ Z}$ vanish is the same as to say the the 3-vectors $\hat{Z} \wedge d_{\hat{P}}\hat{ Z} $  and 
$\hat{Z} \wedge d_{\hat{Q}}\hat{ Z} $ separately vanish. So the first two Lemmas provide an overdetermined set of conditions entailing that the vector field 
$\hat{Z}$ satisfies the Frobenius conditions with respect to the Poisson pencil $\hat{Q}- \lambda \hat{P}$ on the  submanifold $\hat{\cal{S}}$.
The next two Lemmas show how to reduce this overdetermined system of conditions. 

\begin{lemma} The twelve algebraic constraints imposed by the first two Lemmas are equivalent to the four quadratic equations (\ref{quadrics}) presented in the Introduction.
\end{lemma}
{\it Proof.} The first two Eq.s (\ref{quadrics}) are already contained in the set of 12  algebraic constraints. The third equation follows from the obvious identity
\begin{multline*}
\begin{split}
\sum\limits_{\a=1}^3 A_{\a}^{2} + (v+j_{\b}) B_{\g}^2 + (v+j_{\g}) B_{\b}^2 = \sum\limits_{\a=1}^3 (( j_{\b} + j_{\g} ) B_{\a}^{2} + A_{\a}^{2} ) + 2 v \sum\limits_{\a=1}^3 B_{\a}^{2} . \\
\end{split}
\end{multline*}
The fourth equation is proved by noticing that the 12 conditions  imply  
\begin{equation*}
\sum\limits_{\a=1}^3 (v+j_{\b})(v+j_{\g}) B_{\a}^{2} = 0 .
\end{equation*}
By expanding this identity in powers of $v$, and keeping in mind that
\begin{equation*}
v = - \frac{\sum\limits_{\a=1}^3 j_{\a} A_{\a}^{2}}{\sum\limits_{\a=1}^3  A_{\a}^{2}}
\end{equation*}
for the last of the 12 conditions, one readily obtains the fourth equation (\ref{quadrics}). To prove conversely that the twelve equations required by the Lemmas follow from the four quadratic constraints (\ref{quadrics}), 
one simply needs to define the function $v$ according to
\begin{equation*}
v = - \frac{\sum\limits_{\a=1}^3 j_{\a} A_{\a}^{2}}{\sum\limits_{\a=1}^3  A_{\a}^{2}}
\end{equation*}
in such a way to satisfy the last of the 12 conditions. Then one has to keep in mind the parametric representations 
\begin{gather*}
 A_\a = c_{\a} v_{\b} v_{\g},  \quad B_{\a} = c_{\a} v_{\a}
\end{gather*}
of the functions $ A_\a(\zeta)$ and $ B_\a(\zeta) $ introduced in Section 3, and the relation $v_{\a}^2 = v+ j_{\a}$.
With these positions, the 12 algebraic constraints follow immediately. $\square$

\medskip
\begin{lemma}
The six elliptic functions defined by the four quadratic constraints (\ref{quadrics}) solve the three residual ODE's imposed by Lemma \ref{I}
\end{lemma}
{\it Proof.} Let us denote, for brevity, the left-hand side of the differential equations by the symbol $Eq(\a)$, and let us notice that the four quadrics imply the further identity $ \sum\limits_{\a=1}^3 j_{\a} A_{\a} B_{\a} = 0$. Then one easily sees that: 1) The expression $\sum\limits_{\a=1}^3 A_{\a} Eq(\a)$ vanishes on account of the constraint $\sum\limits_{\a=1}^3 A_{\a} B_{\a} = 0 $; 2) The expression $\sum\limits_{\a=1}^3 B_{\a} Eq(\a)$ vanishes on account of the constraints $\sum\limits_{\a=1}^3 A_{\a} B_{\a} = 0 $ and  $\sum\limits_{\a=1}^3 B_{\a}^{2} = 0$; 3) The expression $\sum\limits_{\a=1}^3 j_{\a} B_{\a} Eq(\a)$ vanishes on account of the constraints $\sum\limits_{\a=1}^3 j_{\a} A_{\a} B_{\a} = 0 $ and $\sum\limits_{\a=1}^3 ( j_{\b} + j_{\g} ) B_{\a}^{2} + A_{\a}^{2} = 0$. Since three independent linear combinations of the differential equations vanish on account of the algebraic constraints, one conclude that the equations themselves vanish. This Lemma concludes the proof of Proposition 4. $\square$


\begin{thebibliography}{30}

\bibitem{Clebsch} A. Clebsch.
\"Uber die Bewegung eines K\"orpers in einer Fl\"ussigkeit.
{\it Math. Ann.}, 1871, (3) , 238-261.

\bibitem{Schot} Schottky F.,
\emph{ \"Uber das analytische Problem der Rotation eines starren 
K\"orpers in Raume von vier Dimensionen,} 
Sitzungsber., K\"onig. Preuss. Akad. Wiss., Berlin,  {12} (1891), 227--232.

\bibitem{AvMVh} Adler M., van Moerbeke P. and Vanhaecke P.,
              \emph{Algebraic integrability, Painleve geometry and Lie algebras.}, 
               {Springer},  {(2013)}.
               
\bibitem{avm82}  Adler M., van Moerbeke P.,
                          \emph{ The algebraic integrability  of the geodesic flow on $SO(4)$ ,} 
                            Inventiones Mathematicae, (1982).
               
             
\bibitem{kot892} K\"otter F.,
\emph{Uber die Bewegung eines festen K\"orpers in einerFl\"ussigkeit. I, II,} 
J. reine angew. Math.,  {109} (1892), 51--81, 89--111.

\bibitem{Bel} Belokolos E.D., Bobenko A.I., Enol'sii V.Z., Its A.R., and   Matveev V.B. ,
\emph{Algebro-Geometric Approach to Nonlinear   Integrable Equations,}
 Springer Series in Nonlinear Dynamics,   Springer-Verlag,1994.

\bibitem{ZC} Zhivkov. A, Christov, O.,
\emph{ Effective solutions of Clebsch and C. Neumann systems.}
edoc.hu-berlin.de, (1998).

\bibitem{Skl} Sklyanin E. K.,
              \emph{Separation of variables: New trends},
               {Progr. Theor. Phys.}, {118}, {(1995)}, {35-60}.
               
\bibitem{avm84}  Adler M., van Moerbeke P. ,                           
                              \emph{ Geodesic flow on $so(4)$ and the intersections of quadrics,} 
                            Proc.Natl. Acad. Sci. USA,  {81}, (1984), 4613--4616.
 
\bibitem{hai83} Haine L.,
\emph{ Geodesic flow on $so(4)$ and Abelian surfaces,}
 Math. Ann., {263} (1983), 435--472.

 \bibitem{MaKow1}  Magri F.,
               \emph{The Kowalevski's top and the method of syzygies},
               {Ann. Inst. Fourier}, {55}, {(2005)}, {2146-2159}.

\bibitem{MaKow2}  Magri F.,
              \emph{The Kowalevski top revisited},
               { in Vol. 1 of Integrable sytems and algebraic geometry},
               {  R. Donagi, T. Shaska ed.s,}
               {London Mathematical Society Lecture Notes Series },  {(2020)}, 329-355.
               
               
\bibitem{MaKow3}  Magri F.,
              \emph{The Kowalevski separability conditions},
               { in Dubrovin Memorial volume},{ I. Krichever, S. Novikov, O. Ogievetsky, S. Shlomann ed.s,}
               {Proceedings of Symposia in Pure Mathematics book series}, AMS (to appear).
               
 
\bibitem{Pant} Pantazis S.,\emph{ Prym varieties and the geodesic flow on $SO(n)$,}  Math. Ann., (1986) 273--297.

\bibitem{web878} Weber H.,
              \emph{ Anwendung der Thetafunctionen zweir Veranderlicher auf die Theorie der Bewegung eines festen K\"orpers in einer Fl\"ussigkeit',}
               Math. Ann., {14} (1878), 173--206.

               
 \bibitem{GelZak} Gel'fand I.M.,  Zakharevich I.,
 \emph{On the Local Geomatry of Bihamiltonian Structures,}
 The Gel'fand Mathematical Seminars 1990-1992 ,Birkhauser, Boston, 1993.   

\bibitem{MaSk} Magri F., Skrypnyk T.,
              \emph{The Clebsch system},
              arXiv Math: 1512.04872, {(2015)} .
              

\end{thebibliography}
\end{document}